\documentclass{osa-article}

\journal{oe}

\articletype{Research Article}

\usepackage[group-separator={,}]{siunitx}
\usepackage{graphicx}
\usepackage{amsmath}
\usepackage{xcolor}
\usepackage[normalem]{ulem}

\usepackage[acronym,nomain]{glossaries}
\glsdisablehyper
\newcommand{\SetCapsType}{normalcaps}

\makeatletter
\providecommand{\SetCapsType}{smallcaps}

\long\def\@scTrue{smallcaps}
\long\def\@scFalse{normalcaps}
\newcommand{\acroSCaps}[1]{%
 \begingroup
  \ifx\SetCapsType\@scTrue 
    \textsc{#1}%
  \else
    \MakeUppercase{#1}%
  \fi
  \endgroup
}
\makeatother

\newcommand{\nAcronym}[4][]{%
	\newacronym[#1]{#2}{#3}{#4}
}

\makeatletter
\@ifpackageloaded{babel}{%
    \newcommand{\usuk}[2]{%
        \iflanguage{USenglish}{#1}{#2}%
    }%
}{%
    \newcommand{\usuk}[2]{%
        #1%
    }%
}%
\makeatother

\usepackage[shortcuts]{extdash}
\newcommand{\qam}[1]{
    \ifglsused{QAM}%
        {#1\=/\gls{QAM}}%
        {#1\=/ary \gls{QAM}%
    }%
}%

\nAcronym{2A8PSK}{\acroSCaps{2a8psk}}{2-ary amplitude 8-ary phaseshift keying}

\nAcronym{3CCMCF}{\acroSCaps{3cc-mcf}}{3-core coupled-core multi-core fiber}

\nAcronym{4D}{\acroSCaps{4d}}{four-dimensional}
\nAcronym{4D64PRS}{\acroSCaps{4d-64prs}}{\usuk{four-dimensional 64-ary polarization-ring-switching}{four-dimensional 64-ary polarisation-ring-switching}}

\nAcronym{5B4D2A8PSK}{\acroSCaps{5b4d-2a8psk}}{5-bit four-dimensional two-amplitude 8-ary phase-shift keying}

\nAcronym{6B4D2A8PSK}{\acroSCaps{6b4d-2a8psk}}{6-bit four-dimensional two-amplitude 8-ary phase-shift keying}

\nAcronym{8D}{\acroSCaps{8d}}{eight-dimensional}\nAcronym{8D2048PRS}{\acroSCaps{8d-2048prs}}{eight-dimensional 2048-ary polarization-ring-switching}
\nAcronym{8D2048PRST1}{\acroSCaps{8d-2048prs-t1}}{eight-dimensional 2048-ary polarization-ring-switching type 1}
\nAcronym{8D2048PRST2}{\acroSCaps{8d-2048prs-t2}}{eight-dimensional 2048-ary polarization-ring-switching type 2}

\nAcronym{ABC}{\acroSCaps{abc}}{automatic bias controller}
\nAcronym{ADC}{\acroSCaps{adc}}{analog-to-digital converter}
\nAcronym{AIR}{\acroSCaps{air}}{achievable information rate}
\nAcronym{AOM}{\acroSCaps{aom}}{acoustic optical modulator}
\nAcronym{API}{\acroSCaps{api}}{application programming interface}
\nAcronym{AR}{\acroSCaps{ar}}{achievable rate}
\nAcronym{ASE}{\acroSCaps{ase}}{amplified spontaneous emission}
\nAcronym{ASK}{\acroSCaps{ask}}{amplitude-shift keying}
\nAcronym{ASIC}{\acroSCaps{asic}}{application-specific integrated circuit}
\nAcronym{ARRWG}{\acroSCaps{a}rr\acroSCaps{wg}}{arrayed-waveguide grating}
\nAcronym{AWG}{\acroSCaps{awg}}{arbitrary-waveform generator}
\nAcronym{AWGN}{\acroSCaps{awgn}}{additive white Gaussian noise}

\nAcronym{BCH}{\acroSCaps{bch}}{Bose-Chaudhuri-Hocquenghem}
\nAcronym{BER}{\acroSCaps{ber}}{bit error rate}
\nAcronym{BICM}{\acroSCaps{bicm}}{bit-interleaved coded modulation}
\nAcronym{BMD}{\acroSCaps{bmd}}{bit-metric decoding}
\nAcronym{BPD}{\acroSCaps{bpd}}{balanced photo-diode}
\nAcronym{BPF}{\acroSCaps{bpf}}{bandpass filter}
\nAcronym{BPS}{\acroSCaps{bps}}{blind phase search}
\nAcronym{BPSK}{\acroSCaps{bpsk}}{binary phase-shift keying}
\nAcronym{BRGC}{\acroSCaps{brgc}}{binary reflected Gray code}
\nAcronym{BTB}{\acroSCaps{btb}}{back-to-back}

\nAcronym{CAGR}{\acroSCaps{cagr}}{compound annual growth rate}
\nAcronym{CCDM}{\acroSCaps{ccdm}}{constant composition distribution matching}
\nAcronym{CCF}{\acroSCaps{ccf}}{\usuk{coupled-core fiber}{coupled-core fibre}}
\nAcronym{CD}{\acroSCaps{cd}}{chromatic dispersion}
\nAcronym{CIR}{\acroSCaps{cir}}{channel impulse response}
\nAcronym{CMA}{\acroSCaps{cma}}{constant modulus algorithm}
\nAcronym{CMUX}{\acroSCaps{cmux}}{core multiplexer}
\nAcronym{ChUT}{\acroSCaps{chut}}{channel under test}
\nAcronym{CUT}{\acroSCaps{cut}}{channel under test}
\nAcronym{CPE}{\acroSCaps{cpe}}{carrier phase estimation}
\nAcronym{CPU}{\acroSCaps{cpu}}{central processing unit}
\nAcronym{CSPR}{\acroSCaps{cspr}}{carrier-to-signal power ratio}
\nAcronym{CW}{\acroSCaps{cw}}{continuous wave}

\nAcronym{DA}{\acroSCaps{da}}{driver amplifier}
\nAcronym{DAC}{\acroSCaps{dac}}{digital-to-analog converter}
\nAcronym{DCF}{\acroSCaps{dcf}}{\usuk{dispersion compensated fiber}{dispersion compensated fibre}}
\nAcronym{DCI}{\acroSCaps{dci}}{data center interconnect}
\nAcronym{DDLMS}{\acroSCaps{dd-lms}}{decision-directed least mean square}
\nAcronym{DFB}{\acroSCaps{dfb}}{distributed feedback}
\nAcronym{DGD}{\acroSCaps{dgd}}{differential group delay}
\nAcronym{DH}{\acroSCaps{dh}}{digital holography}
\nAcronym{DM}{\acroSCaps{dm}}{distribution matcher}
\nAcronym{DMA}{\acroSCaps{dma}}{direct memory access}
\nAcronym{DMD}{\acroSCaps{dmd}}{differential mode delay}
\nAcronym{DMG}{\acroSCaps{dmg}}{differential modal gain}
\nAcronym{DMGD}{\acroSCaps{dmgd}}{differential mode group delay}
\nAcronym{DML}{\acroSCaps{dml}}{directly-modulated laser}
\nAcronym{DP}{\acroSCaps{dp}}{\usuk{dual-polarization}{dual-polarisation}}
\nAcronym{DPC}{\acroSCaps{dpc}}{digital pre-compensation}
\nAcronym{DPE}{\acroSCaps{dpe}}{digital pre-emphasis}
\nAcronym{DPIQ}{\acroSCaps{dp-iqm}}{\usuk{dual-polarization IQ-modulator}{dual-polarisation IQ-modulator}}
\nAcronym{DPLL}{\acroSCaps{dpll}}{digital phase-locked loop}
\nAcronym{DQPSK}{\acroSCaps{dqpsk}}{differential quaternary phase-shift-keying}
\nAcronym{DRA}{\acroSCaps{dra}}{distributed Raman amplifier}
\nAcronym{DRE}{\acroSCaps{dre}}{digital resolution enhancer}
\nAcronym{DSF}{\acroSCaps{dsf}}{\usuk{dispersion-shifted fiber}{dispersion-shifted fibre}}
\nAcronym{DSO}{\acroSCaps{dso}}{digital sampling oscilloscope}
\nAcronym{DSP}{\acroSCaps{dsp}}{digital signal processing}
\nAcronym{DUT}{\acroSCaps{dut}}{device-under-test}
\nAcronym{DWDM}{\acroSCaps{dwdm}}{dense wavelength-division multiplexing}

\nAcronym{EAM}{\acroSCaps{eam}}{electro-absorption modulator}
\nAcronym{ECL}{\acroSCaps{ecl}}{external cavity laser}
\nAcronym{ED}{\acroSCaps{ed}}{Eucledian distance}
\nAcronym{EDFA}{\acroSCaps{edfa}}{\usuk{erbium-doped fiber amplifier}{erbium-doped fibre amplifier}}
\nAcronym{ENOB}{\acroSCaps{enob}}{effective number of bits}
\nAcronym{ESS}{\acroSCaps{ess}}{enumerative sphere shaping}

\nAcronym{FD}{\acroSCaps{fd}}{frequency domain}
\nAcronym{FDE}{\acroSCaps{fde}}{\usuk{frequency domain equalizer}{frequency domain equaliser}}
\nAcronym{FEC}{\acroSCaps{fec}}{forward error correction}
\nAcronym{FFT}{\acroSCaps{fft}}{fast Fourier transform}
\nAcronym{FIR}{\acroSCaps{fir}}{finite impulse response}
\nAcronym{FMF}{\acroSCaps{fmf}}{\usuk{few-mode fiber}{few-mode fibre}}
\nAcronym[plural=FM-MCF, firstplural=\usuk{few-mode multi-core fibers}{few-mode multi-core fibres}]{FM-MCF}{\acroSCaps{fm-mcf}}{\usuk{few-mode multi-core fiber}{few-mode multi-core fibre}}
\nAcronym{FPGA}{\acroSCaps{fpga}}{field-programmable gate array}
\nAcronym{FWM}{\acroSCaps{fwm}}{four-wave mixing}

\nAcronym{GD}{\acroSCaps{gd}}{group delay}
\nAcronym{GI}{\acroSCaps{gi}}{graded-index}
\nAcronym{GFF}{\acroSCaps{gff}}{gain flattening filter}
\nAcronym{GIMMF}{\acroSCaps{gi-mmf}}{\usuk{graded-index multi-mode fiber}{graded-index multi-mode fibre}}
\nAcronym{GMI}{\acroSCaps{gmi}}{\usuk{generalized mutual information}{generalised mutual information}}
\nAcronym{GNSE}{\acroSCaps{gnse}}{generalized nonlinear Schr\"{o}dinger equation}
\nAcronym{GPU}{\acroSCaps{gpu}}{graphics processing unit}
\nAcronym{GS}{\acroSCaps{gs}}{geometric shaping}
\nAcronym{GV}{\acroSCaps{gv}}{group-velocity}
\nAcronym{GVD}{\acroSCaps{gvd}}{group-velocity dispersion}
\nAcronym{GPIO}{\acroSCaps{gpio}}{general purpose input output}

\nAcronym{HDFEC}{\acroSCaps{hd-fec}}{hard decision forward error correction}

\nAcronym[plural=IL, firstplural=insertion losses (\textsc{il})]{IL}{\acroSCaps{il}}{insertion loss}
\nAcronym{IFFT}{\acroSCaps{ifft}}{inverse fast Fourier transform}
\nAcronym{IMDD}{\acroSCaps{im/dd}}{intensity-modulation direct-detection}
\nAcronym{IQM}{\acroSCaps{iqm}}{in-phase and quadrature modulator}
\nAcronym{ISI}{\acroSCaps{isi}}{intersymbol interference}

\nAcronym{KK}{\acroSCaps{kk}}{Kramers-Kronig}

\nAcronym{LCOS}{\acroSCaps{LCoS}}{liquid crystal on silicon}
\nAcronym{LDPC}{\acroSCaps{ldpc}}{low-density parity-check}
\nAcronym{LEAF}{\acroSCaps{leaf}}{\usuk{large effective area fiber}{large effective area fibre}}
\nAcronym{LFSR}{\acroSCaps{lfsr}}{linear-feedback shift register}
\nAcronym{LMS}{\acroSCaps{lms}}{least means square}
\nAcronym{LLR}{\acroSCaps{llr}}{log-likelihood ratio}
\nAcronym{LO}{\acroSCaps{lo}}{local oscillator}
\nAcronym{LP}{\acroSCaps{lp}}{\usuk{linearly polarized}{linearly polarised}}
\nAcronym{LSPS}{\acroSCaps{lsps}}{\usuk{loop-synchronized polarization scrambler}{loop-synchronised polarisation scrambler}}
\nAcronym{LUT}{\acroSCaps{lut}}{lookup table}

\nAcronym{MB}{\acroSCaps{mb}}{Maxwell-Bolzmann}
\nAcronym{MCF}{\acroSCaps{mcf}}{\usuk{multi-core fiber}{multi-core fibre}}
\nAcronym{MDG}{\acroSCaps{mdg}}{mode dependent gain}
\nAcronym[plural=MDL, firstplural=mode-dependent losses (\textsc{mdl})]{MDL}{\acroSCaps{mdl}}{mode-dependent loss}
\nAcronym{MDM}{\acroSCaps{mdm}}{mode division multiplexing}
\nAcronym{MEMS}{\acroSCaps{mems}}{micro-electro-mechanical systems}
\nAcronym{MF}{\acroSCaps{mf}}{matched filter}
\nAcronym{MI}{\acroSCaps{mi}}{mutual information}
\nAcronym{MIMO}{\acroSCaps{mimo}}{multiple-input multiple-output}
\nAcronym{ML}{\acroSCaps{ml}}{machine learning}
\nAcronym{MMA}{\acroSCaps{mma}}{multi-modulus algorithm}
\nAcronym{MMF}{\acroSCaps{mmf}}{\usuk{multi-mode fiber}{multi-mode fibre}}
\nAcronym{MMSE}{\acroSCaps{mmse}}{minimum mean squared error}
\nAcronym{MP}{\acroSCaps{mp}}{minimum phase}
\nAcronym{MPLC}{\acroSCaps{mplc}}{multi-plane light converter}
\nAcronym{MSE}{\acroSCaps{mse}}{mean squared error}
\nAcronym{MUX}{\acroSCaps{mux}}{multiplexer}
\nAcronym{MZM}{\acroSCaps{mzm}}{Mach-Zehnder modulator}
\nAcronym{MZI}{\acroSCaps{mzi}}{Mach-Zehnder interferometer}

\nAcronym{NA}{\acroSCaps{na}}{numerical aperture}
\nAcronym{NF}{\acroSCaps{nf}}{noise figure}
\nAcronym{NGMI}{\acroSCaps{ngmi}}{\usuk{normalized generalized mutual information}{normalised generalised mutual information}}
\nAcronym{NLSE}{\acroSCaps{nlse}}{nonlinear Schr\"{o}ding equation}
\nAcronym{NIR}{\acroSCaps{nir}}{near-infrared}

\nAcronym{OBTB}{\acroSCaps{obtb}}{optical back-to-back}
\nAcronym{ODE}{\acroSCaps{ode}}{ordinary differential equation}
\nAcronym{ODL}{\acroSCaps{odl}}{optical delay line}
\nAcronym{OFC}{\acroSCaps{ofc}}{Optical Fiber Communications Conference}
\nAcronym{OFDR}{\acroSCaps{ofdr}}{optical frequency-domain reflectometer}
\nAcronym{OFDM}{\acroSCaps{ofdm}}{orthogonal frequency division multiplexing}
\nAcronym{OMFT}{\acroSCaps{omft}}{optical multi-format transmitter}
\nAcronym{OOK}{\acroSCaps{ook}}{on-off keying}
\nAcronym{OPLL}{\acroSCaps{opll}}{optical phase-locked loop}
\nAcronym{OSA}{\acroSCaps{osa}}{\usuk{optical spectrum analyzer}{optical spectrum analyser}}
\nAcronym{OSNR}{\acroSCaps{osnr}}{optical signal-to-noise ratio}
\nAcronym{OTDR}{\acroSCaps{otdr}}{optical time-domain reflectometer}
\nAcronym{OTF}{\acroSCaps{otf}}{optical tunable filter}
\nAcronym{OVNA}{\acroSCaps{ovna}}{\usuk{optical vector network analyzer}{optical vector network analyser}}

\nAcronym{PAM}{\acroSCaps{pam}}{pulse-amplitude modulation}
\nAcronym{PAS}{\acroSCaps{pas}}{probabilistic amplitude shaping}
\nAcronym{PAPR}{\acroSCaps{papr}}{peak-to-avarage power ratio}
\nAcronym{PBC}{\acroSCaps{pbc}}{\usuk{polarization beam combiner}{polarisation beam combiner}}
\nAcronym{PBS}{\acroSCaps{pbs}}{polarization beam splitter}
\nAcronym{PCVD}{\acroSCaps{pcvd}}{plasma chemical vapor depostion}
\nAcronym{PD}{\acroSCaps{pd}}{photodiode}
\nAcronym{PDL}{\acroSCaps{pdl}}{polarization-dependent loss}
\nAcronym{PDM}{\acroSCaps{pdm}}{\usuk{polarization-division multiplexing}{polarisation-division multiplexing}}
\nAcronym{PL}{\acroSCaps{pl}}{photonic lantern}
\nAcronym{PMBPSK}{\acroSCaps{pm-bpsk}}{polarization-multiplexed binary phase-shift-keying}
\nAcronym{PMQPSK}{\acroSCaps{pm-qpsk}}{polarization-multiplexed quaternary phase-shift-keying}
\nAcronym{PM8QAM}{\acroSCaps{pm-8qam}}{polarization-multiplexed 8-ary quadrature amplitude modulation}
\nAcronym{PMD}{\acroSCaps{pmd}}{polarization mode dispersion}
\nAcronym{PNOB}{\acroSCaps{pnob}}{physical number of bits}
\nAcronym{PON}{\acroSCaps{pon}}{passive-optical network}
\nAcronym{PRBS}{\acroSCaps{prbs}}{pseudorandom bit sequence}
\nAcronym{PS}{\acroSCaps{ps}}{probabilistic shaping}
\nAcronym{PSK}{\acroSCaps{psk}}{phase-shift keying}
\nAcronym{PSP}{\acroSCaps{psp}}{\usuk{principal states of polarization}{principal states of polarisation}}

\nAcronym{QAM}{\acroSCaps{qam}}{quadrature amplitude modulation}
\nAcronym{QPSK}{\acroSCaps{qpsk}}{quaternary phase-shift-keying}

\nAcronym{RAM}{\acroSCaps{ram}}{random-access memory}
\nAcronym{RF}{\acroSCaps{rf}}{radio frequency}
\nAcronym{RRC}{\acroSCaps{rrc}}{root-raised-cosine}
\nAcronym{ROADM}{\acroSCaps{roadm}}{reconfigurable optical add-drop multiplexer}
\nAcronym{ROI}{\acroSCaps{roi}}{region of interest}

\nAcronym{SA}{\acroSCaps{sa}}{simulated annealing}
\nAcronym{SamPerSym}{\acroSCaps{sps}}{samples per symbol}
\nAcronym{SBS}{\acroSCaps{sbs}}{stimulated Brillouin scattering}
\nAcronym{SDFEC}{\acroSCaps{sd-fec}}{soft decision forward error correction}
\nAcronym{SDM}{\acroSCaps{sdm}}{space-division multiplexing}
\nAcronym{SE}{\acroSCaps{se}}{spectral efficiency}
\nAcronym{SER}{\acroSCaps{ser}}{symbol error rate}
\nAcronym{SI}{\acroSCaps{si}}{step index}
\nAcronym{SLM}{\acroSCaps{slm}}{spatial light modulator}
\nAcronym{SMF}{\acroSCaps{smf}}{\usuk{single-mode fiber}{single-mode fibre}}
\nAcronym{SNR}{\acroSCaps{snr}}{signal-to-noise ratio}
\nAcronym{SOA}{\acroSCaps{soa}}{semiconductor optical amplifier}
\nAcronym[firstplural=\usuk{states of polarization (\acroSCaps{sop})}{states of polarisation (\acroSCaps{sop})}]{SOP}{\acroSCaps{sop}}{\usuk{state of polarization}{state of polarisation}}
\nAcronym{SPM}{\acroSCaps{spm}}{self-phase modulation}
\nAcronym{SPS}{\acroSCaps{sps}}{\usuk{synchronized polarization scrambler}{synchronised polarisation scrambler}}
\nAcronym{SRS}{\acroSCaps{srs}}{stimulated Raman scattering}
\nAcronym{SSBI}{\acroSCaps{ssbi}}{signal-signal beat interference}
\nAcronym{SSFM}{\acroSCaps{ssfm}}{split-step Fourier method}
\nAcronym{SSMF}{\acroSCaps{ssmf}}{\usuk{standard single-mode fiber}{standard single-mode fibre}}
\nAcronym{STL}{\acroSCaps{stl}}{swept tunable laser}
\nAcronym{SVD}{\acroSCaps{svd}}{singular value decomposition}
\nAcronym{SWI}{\acroSCaps{swi}}{swept wavelength interferometry}

\nAcronym{TD}{\acroSCaps{td}}{time domain}
\nAcronym{TDE}{\acroSCaps{tde}}{\usuk{time domain equalizer}{time domain equaliser}}
\nAcronym{TDMSDM}{\acroSCaps{tdm-sdm}}{time-domain multiplexed space-division multiplexing}
\nAcronym{TE}{\acroSCaps{te}}{transverse electric}
\nAcronym{TIA}{\acroSCaps{tia}}{trans-impedance amplifier}
\nAcronym{TM}{\acroSCaps{TM}}{transverse magnetic}
\nAcronym{TH4D}{\acroSCaps{th-4d}}{time domain hybrid four-dimensional}
\nAcronym{TH4D2A8PSK}{\acroSCaps{th-4d-2a8psk}}{time domain hybrid four-dimensional two-amplitude eight-phase shift keying}

\nAcronym{VCSEL}{\acroSCaps{vcsel}}{vertical-cavity surface emitting laser}
\nAcronym{VOA}{\acroSCaps{voa}}{variable optical attenuator}

\nAcronym{WDM}{\acroSCaps{wdm}}{wavelength-division multiplexing}
\nAcronym{WGA}{\acroSCaps{wga}}{weakly guiding approximation}
\nAcronym{WGN}{\acroSCaps{wgn}}{white Gaussian noise}
\nAcronym{WSS}{\acroSCaps{wss}}{wavelength selective switch}

\nAcronym{XPM}{\acroSCaps{xpm}}{cross-phase modulation}
\nAcronym{XT}{\acroSCaps{xt}}{cross-talk}

\nAcronym{3DWG}{\acroSCaps{3dwg}}{3D-waveguide}

\usepackage[style=ieee, defernumbers=true, dashed=false]{biblatex}
\bibliography{ref} 

\DeclareBibliographyCategory{ignore}

\usepackage{lipsum}

\usepackage{caption}
\usepackage{subcaption}

\usepackage[capitalise]{cleveref}

\makeatletter
\let\blx@rerun@biber\relax
\makeatother

\begin{document}

\title{Real-time Transmission of Geometrically-shaped Signals using a Software-defined GPU-based Optical Receiver}

\author{Sjoerd~van~der~Heide\authormark{1,*},
        Ruben~S.~Luis\authormark{2},
        Sebastiaan~Goossens\authormark{1},
        Benjamin~J.~Puttnam\authormark{2},
        Georg~Rademacher\authormark{2},
        Ton~Koonen\authormark{1},
        Satoshi~Shinada\authormark{2},
        Yoshinari~Awaji\authormark{2},
        Alex~Alvarado\authormark{1},
        Hideaki~Furukawa\authormark{2},
        Chigo~Okonkwo\authormark{1}}

\address{\authormark{1}Department of Electrical Engineering, Eindhoven University of Technology, PO Box 513, 5600 MB, Eindhoven, the Netherlands\\
\authormark{2}National Institute of Information and Communications Technology, Photonic Network System Laboratory, 4-2-1, Nukui-Kitamachi, Koganei, Tokyo, 184-8795,  Japan}

\email{\authormark{*}s.p.v.d.heide@tue.nl} 



\begin{abstract}
A software-defined optical receiver is implemented on an off-the-shelf commercial graphics processing unit (GPU). The receiver provides real-time signal processing functionality to process 1~GBaud minimum phase (MP) 4-, 8-, 16-, 32-, 64-, 128-ary quadrature amplitude modulation (QAM) as well as geometrically shaped (GS) 8- and 128-QAM signals using Kramers-Kronig (KK) coherent detection. Experimental validation of this receiver over a 91~km field-deployed optical fiber link between two Tokyo locations is shown with detailed optical signal-to-noise ratio (OSNR) investigations. A net data rate of 5~Gbps using 64-QAM is demonstrated.
\end{abstract}

\section{Introduction}
\label{sec:introduction}

In recent years, \glspl{GPU} have been proposed as an alternative to \glspl{FPGA} \cite{Beppu_FPGA_2020,Beppu_2_FPGA_2020,Randel_FPGA_2015} and \glspl{ASIC} for optical communications \cite{Li_realtime_LDPC, suzuki_fec,suzuki_phy_2018, suzuki_phy_2019,suzuki_phy_2020,suzuki_real-time_2020,ECOC_realtimeGPU,gpuJLT,OFC_realtime10k}. More than a decade of steady exponential improvement of computation capacity (45\% yearly increase \cite{winzer_scaling_2017}) and energy efficiency (25\% yearly increase \cite{sun_summarizing_2019}) of \glspl{GPU} have accelerated its potential applications in the field of optical communications. Recent demonstrations include real-time \gls{FEC} decoding \cite{Li_realtime_LDPC, suzuki_fec}, physical-layer functionality \cite{suzuki_phy_2018, suzuki_phy_2019, suzuki_phy_2020}, \gls{DQPSK} detection \cite{suzuki_real-time_2020}, and flexible multi-modulation format detection using directly detected pulse-amplitude modulated signals, and coherently detected \gls{QAM} signals \cite{ECOC_realtimeGPU,gpuJLT,OFC_realtime10k} using \gls{KK} detection \cite{mecozzi_kramers_2016}. We demonstrated a real-time receiver over a \SI{10000}{km} straight-line link \cite{OFC_realtime10k} and a field-deployed fiber \cite{gpuJLT}.
 
In this work, the flexible, software-defined real-time multi-modulation format receiver is optimized for improved performance and demonstrated modulation formats are extended with odd-power \gls{QAM} and \gls{GS}. A commercial off-the-shelf \gls{GPU} is used for real-time digital signal processing of \gls{MP} \qam{4-, 8-, 16-, 32-, 64-, and 128} as well as geometrically-shaped \qam{8} and \qam{128} signals using \gls{KK} coherent detection \cite{mecozzi_kramers_2016}. The receiver is experimentally validated using a field-deployed transmission link between two Tokyo locations with dynamic components of the receiver \gls{DSP} handling fluctuations of the environmental conditions. Detailed investigations into \gls{OSNR} performance are shown and \gls{CSPR} is optimized for all transmission scenarios. Net throughput is calculated for the evaluated formats and multiple \gls{HDFEC} algorithms, with \qam{64} paired with a 20\% overhead \gls{HDFEC} achieving a net throughput of \SI{5}{Gbps} at \SI{28.2}{dB} \gls{OSNR}. 

This paper is an extension to \cite{gpuJLT}. The demonstrated capabilities of the real-time receiver are extended to show support for odd-power \gls{QAM} and geometrically-shaped constellations. Furthermore, the performance of the receiver with respect to \cite{gpuJLT} is improved through several optimizations; Firstly, the power of the digitally-inserted carrier tone with respect to the signal (\gls{CSPR}) is optimized. In \cite{gpuJLT}, a static \gls{CSPR} value was used, here, a detailed investigation into the influence of \gls{CSPR} on transmission performance is presented. Secondly, the \glspl{ROADM} are removed from the transmission path and replaced by an \gls{EDFA} at the remote site, enabling evaluation of the receiver at higher \gls{OSNR}. Thirdly, the \SI{1}{GHz} photodiode is replaced by a \SI{6.5}{GHz} model, significantly improving the receiver bandwidth. Fourthly, the gap between the signal and digitally-inserted carrier is optimized, resulting in the use of a smaller gap. Fifthly, launch power is optimized, but we observe no significant launch power dependence on \gls{OSNR} performance. These changes and optimizations substantially improve transmission performance compared to what we presented in \cite{gpuJLT}. Finally, more modulation formats are evaluated, \qam{4-, 8-, GS-8-, 16-, 32-, 64-, 128-, and GS-128} to comprehensively demonstrate multi-modulation format receiver capability.


\begin{figure*}
\includegraphics[width=\linewidth]{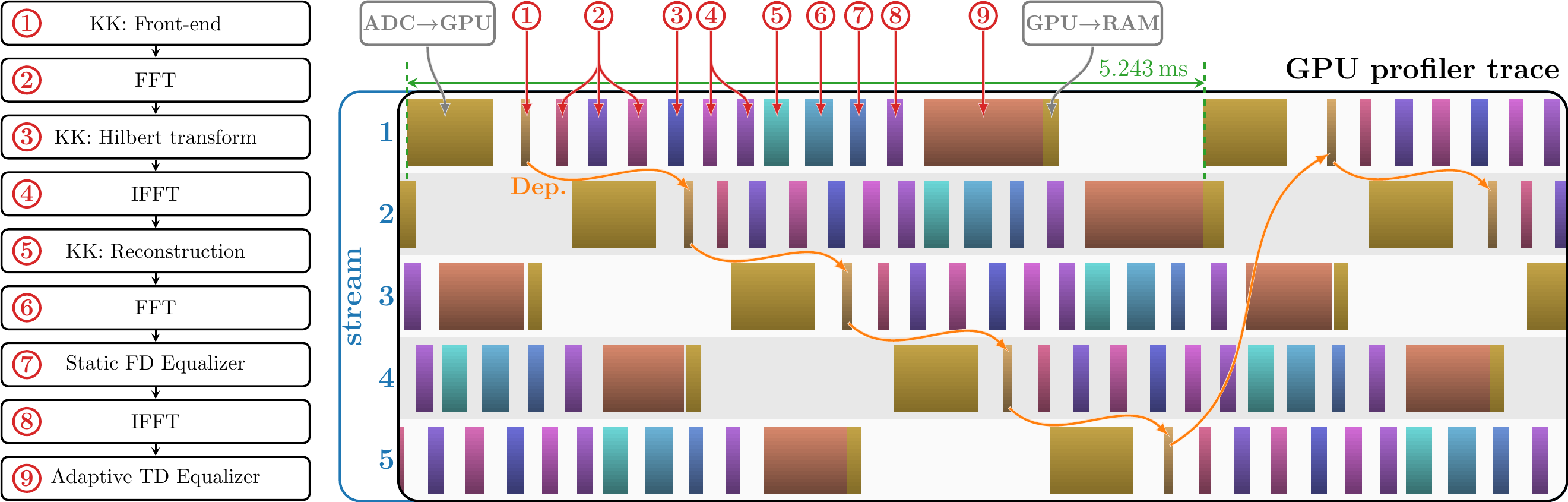}
\caption{Real-time Kramers-Kronig digital signal processing chain including GPU profiler trace, additional details in \cite{gpuJLT}.}
\label{fig:kkdsp} 
\end{figure*}

\section{Digital signal processing chain}
\label{sec:dsp}

\cref{fig:kkdsp} shows the real-time \gls{DSP} chain for \gls{KK} coherent \qam{N} signals. Here, an overview of its GPU-based implementation is given. A more detailed description can be found in \cite{gpuJLT}. First, buffers containing 2\textsuperscript{22} digitized samples are transferred from \gls{ADC} to \gls{GPU} in real time using \gls{DMA}. The signal processing starts with a \gls{GPU} kernel converting samples  received as  12-bit fixed point to 32-bit floating point numbers, adding the appropriate DC offset, and performing the square root and logarithm \gls{KK} front-end operations. This first kernel is annotated by the number 1 in \cref{fig:kkdsp}. In step 3, enabled by a pair of 100\% overlap-save 1024-point \glspl{FFT}, the phase of the optical signal is recovered by a frequency-domain Hilbert transform. This phase is combined with the amplitude calculated in step 1 to reconstruct the optical signal \cite{mecozzi_kramers_2016} which is subsequently downconverted for further processing. Another pair of \glspl{FFT} supports frequency-domain static equalization and resampling from 4 to 2 samples-per-symbol. Finally, a 4-tap adaptive time-domain widely-linear \cite{Silva_WidelyLinear} \gls{DDLMS} equalizer is employed to recover the signal. The minimum Euclidean distance decisions made by the equalizer are demapped into bits and sent to \gls{RAM}.

Full utilization of \gls{GPU} resources is achieved through massive parallelization within kernels as well as operating multiple processing streams in parallel. Most of the kernels operating within each stream are highly parallel themselves, e.g. the \glspl{FFT}, Hilbert transform, and the frequency-domain equalizer. However, algorithms such as the adaptive time-domain equalizer are hard to parallelize due to its sequential nature and time-dependencies of the tap updates. The use of multiple processing streams allows these hard-to-parallelize algorithms to run next to easy-to-parallelize algorithms. Therefore, even though the adaptive equalization takes up significant amount of \textit{time} as shown in the \gls{GPU} profiler trace in \cref{fig:kkdsp},  significant amount of \textit{resources} is not required. This concept and many other implementation strategies are discussed in detail in \cite{gpuJLT}. Constellation cardinality only slightly influences \gls{GPU} resource utilization. The \gls{KK} algorithm and static equalization take up the majority of computational resources.

Note that several parameters of the real-time receiver are static and optimized offline. Since the \gls{ADC} is AC-coupled, the DC-term of the signal is lost. This DC-term is crucial for correct signal reconstruction and is added in step 1 as described above. The question remains how to determine the optimal value since it is dependent on the signal power, noise power, and \gls{CSPR}. Throughout this work, all measurements are performed multiple times using different DC offset values to ensure the optimal performance is observed. Alternatively, one could implement an algorithm to calculate and update the optimal DC-term in real time \cite{Luis:20}.

Similarly, the 203-tap static frequency-domain equalizer is optimized offline using a training sequence every time that the data aquisition is initialized. Therefore, the entire signal processing chain up to the adaptive equalizer is agnostic to the modulation format. The adaptive equalizer only needs to know the constellation since, after initial setup and convergence using a training sequence, it is updated in a blind decision-directed fashion where part of a buffer is used to update equalizer taps for subsequent buffers. Also, the symbol decisions are demapped into bits and are considered as the output of the \gls{DSP} chain. The constellation points and bit mapping are uploaded to the \gls{GPU} for the equalizer to make decisions based on a minimum Euclidean distance criterion. Note that no phase compensation algorithm is required since the \gls{KK} coherent receiver scheme is free of phase noise. Furthermore, the \gls{DSP} chain does not rely on specific properties of modulation formats such as symmetry. Therefore, support for various geometrically-shaped constellations is achieved by uploading the location of the constellation points and corresponding bit mapping for the equalizer and demapper to use.

\begin{figure*}
\includegraphics[width=\linewidth]{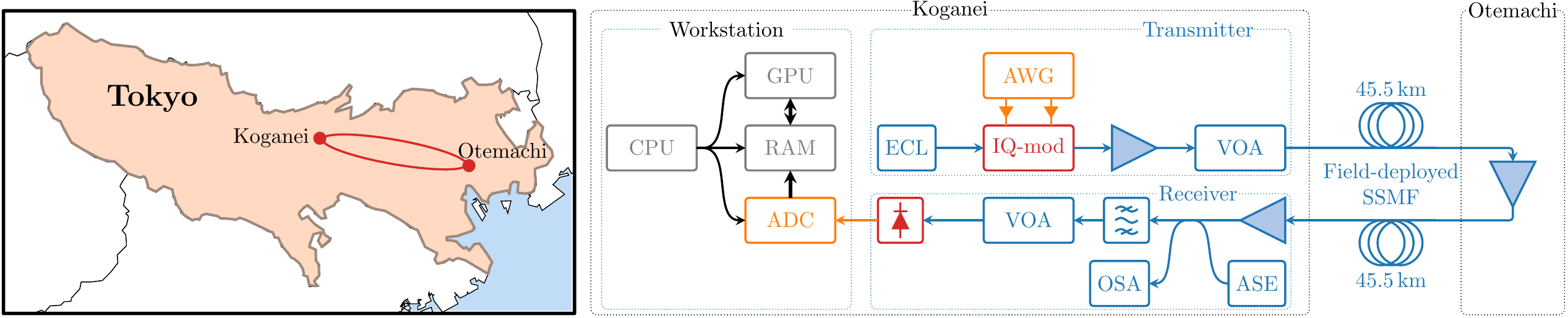}
\caption{Experimental setup with a field-deployed link between Koganei and Otemachi, Tokyo, Japan.}
\label{fig:expsetup} 
\end{figure*}

\section{Experimental Setup}
\label{sec:expsetup}

\cref{fig:expsetup} shows the experimental setup using the field-deployed link between Koganei and Otemachi, Tokyo. First, an 8-bit \SI{12}{GS/s} \gls{AWG} generates \gls{MP} \SI{1}{GBaud} \gls{QAM} signals with 1\% rolloff \gls{RRC} pulse shaping with a digitally inserted carrier tone at \SI{0.516}{GHz}. The 2\textsuperscript{20} N-ary symbol sequence is generated using PCG64 pseudo-random numbers and mapped to the desired modulation format. Both conventional and geometrically shaped \gls{QAM} formats are employed, the latter optimized for the \gls{AWGN} channel and generated through an iterative optimization process similar to \cite{bin_chen_gs_2018}. Further details can be found in \cref{sec:discussion}.

A \SI{16}{GHz} \gls{IQM} modulates the signal onto a \SI{100}{kHz} linewidth \SI{1550.51}{nm} \gls{ECL}. The amplified signal is launched into \SI{45.5}{km} of field-deployed \gls{SSMF} at a fixed launch power. At the other location in Otemachi, Tokyo, an \gls{EDFA} amplifies the signal again to the same launch power and it is transmitted back to Koganei via \SI{45.5}{km} of \gls{SSMF}. 56\% of this fiber is installed in underground ducts and the remainder on aerial paths and in the surface along railway tracks.

At the receiver, the signal is amplified, combined with \gls{ASE} from a noise-loading stage to vary the \gls{OSNR}, and filtered using a \SI{5}{GHz} \gls{BPF}. A \gls{VOA} controls the optical power into the \SI{6.5}{GHz} photodiode such that the electrical output swing fills the \SI{1}{GHz} \SI{4}{GS/s}  12-bit \gls{ADC} detection range. Receiver clock is synchronized to the transmitter and \gls{DSP} is performed in real time on the \gls{GPU}  with 5120 processing cores as described in \cref{sec:dsp}. Error counting is performed offline over 98 buffers containing 2\textsuperscript{20} symbols each.

Several optimizations have been employed to increase \gls{OSNR} performance. Firstly, no significant dependence of launch power on \gls{OSNR} performance was observed. Therefore, the launch power was fixed. Next, the gap between the digitally-inserted carrier required for \gls{KK} detection was optimized. With the 1\% roll-off \gls{RRC} \SI{1}{GBaud} signal ending at \SI{0.505}{GHz} and the carrier tone at \SI{0.516}{GHz}, a gap of only \SI{11}{MHz} was left. One would expect a larger gap to be more beneficial since it reduces \gls{SSBI} which when not entirely removed by the \gls{KK} algorithm leaves reconstruction errors. However, the \gls{ADC} has a limited \SI{3}{dB} bandwidth of \SI{1}{GHz}, thus impairing the signal after detection but crucially before the \gls{KK} algorithm, leading to imperfect reconstruction. It is expected that the penalty from increased imperfect reconstruction obstructs the use of a larger gap. Finally, the power of the digitally-inserted carrier tone relative to the signal, the \gls{CSPR}, is varied and reported in the next section.

\section{Results}
\label{sec:results}

\begin{figure*}
\includegraphics[width=\linewidth]{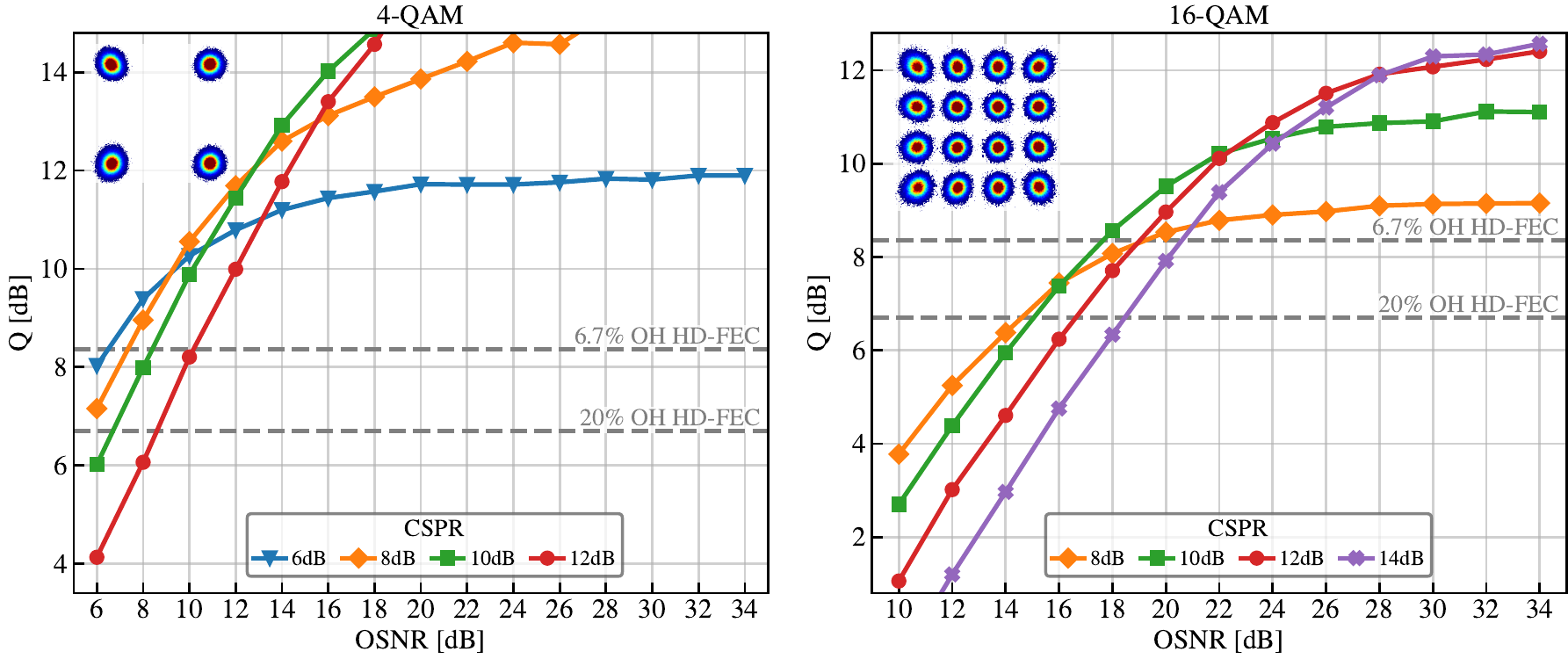}
\caption{Q-factor versus OSNR after transmission through \SI{91}{km} of field-deployed SSMF for 4-QAM and 16-QAM for various CSPRs. Insets show constellations with highest Q-factor without noise loading with \SI{12}{dB} \gls{CSPR} for 4-QAM and \SI{14}{dB} for 16-QAM.}
\label{fig:osnr0} 
\end{figure*}

\cref{fig:osnr0} shows the Q-factor versus \gls{OSNR} for \qam{4} and \qam{16} for multiple \glspl{CSPR}. It is clear that lower \gls{CSPR} values perform better in the low \gls{OSNR} regime and higher \glspl{CSPR} at higher \gls{OSNR}. Note that total signal power as defined in the \gls{OSNR} includes both the carrier tone and the \qam{N} signal. Therefore, if \gls{CSPR} is increased, signal power decreases, \gls{SNR} decreases, and Q-factor decreases. This is especially relevant at low \glspl{OSNR}, because optical noise is dominant in this regime. Conversely, if \gls{CSPR} is increased, carrier power increases, \gls{SSBI} decreases, fewer reconstruction errors occur, and Q-factor increases. This is relevant at higher \glspl{OSNR}, because reconstruction errors are dominant in this regime. Reconstruction errors originate from minimum-phase condition violations and predominantly distort the outer points of a constellation as can be seen in the insets of \cref{fig:osnr0} \cite{mecozzi_kramers_2016}. The \gls{CSPR} trade-off between noise and reconstruction errors is observed for all tested modulations. 

\qam{4} with \SI{6}{dB} \gls{CSPR} reaches the 6.7\% overhead \gls{HDFEC} Q-factor threshold of \SI{8.35}{dB} \cite{7pctFEC} at \SI{6.5}{dB} \gls{OSNR}. For this specific \gls{CSPR}, the Q-factor never exceeds \SI{12}{dB} whilst higher \glspl{CSPR} have a lower error floor or Q-factor ceiling. The optimal \gls{CSPR} depends on the operating regime of the transmission link, the modulation format, and the \gls{FEC} algorithm employed. Q-values above \SI{15}{dB} are not displayed in \cref{fig:osnr0} because too few errors were recorded for statistical significance. \qam{16} reaches the 20\% overhead \gls{HDFEC} Q-factor threshold of \SI{6.70}{dB} \cite{20pctFEC1,20pctFEC2} at \SI{14.6}{dB} \gls{OSNR} using a \gls{CSPR} of \SI{8}{dB}. The 6.7\% overhead \gls{HDFEC} is reached at an \gls{OSNR} of \SI{17.6}{dB} using \SI{10}{dB} \gls{CSPR}.

\begin{figure*}
\includegraphics[width=\linewidth]{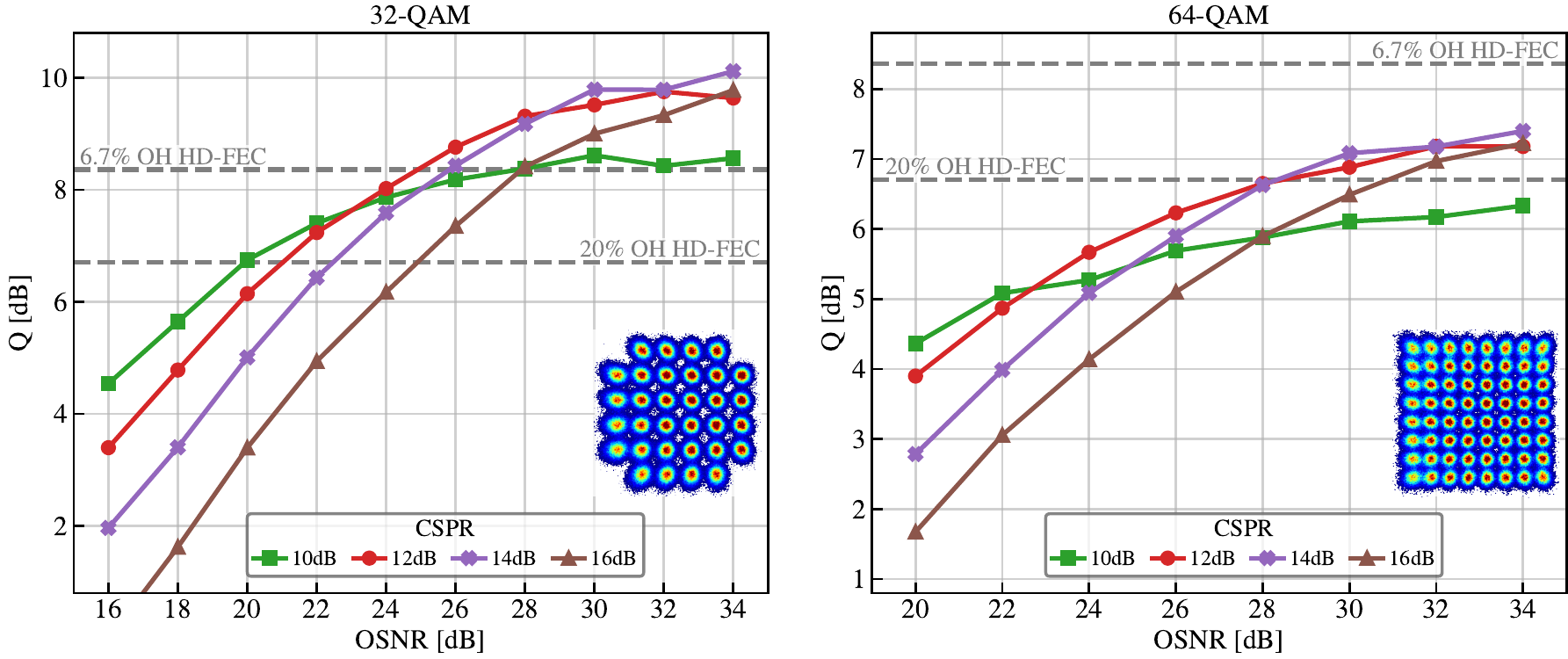}
\caption{Q-factor versus OSNR after transmission through \SI{91}{km} of field-deployed SSMF  for 32-QAM and 64-QAM for various CSPRs. Insets show constellations with highest Q-factor without noise loading with \SI{16}{dB} \gls{CSPR}.}
\label{fig:osnr1} 
\end{figure*}

\gls{OSNR} performance for \qam{32} and \qam{64} are shown in \cref{fig:osnr1}. \qam{32} is able to reach both the 20\% and 6.7\% overhead \gls{FEC} limit at \SIlist{19.9;24.9}{dB}, respectively. However, \qam{64} is not able to reach the 6.7\% overhead \gls{FEC} threshold, but reaches the other threshold at \SI{28.2}{dB}. The successful transmission of \qam{64} is enabled by improving receiver bandwidth, increasing \gls{OSNR}, and optimizing \gls{CSPR}.

\begin{figure*}
\includegraphics[width=\linewidth]{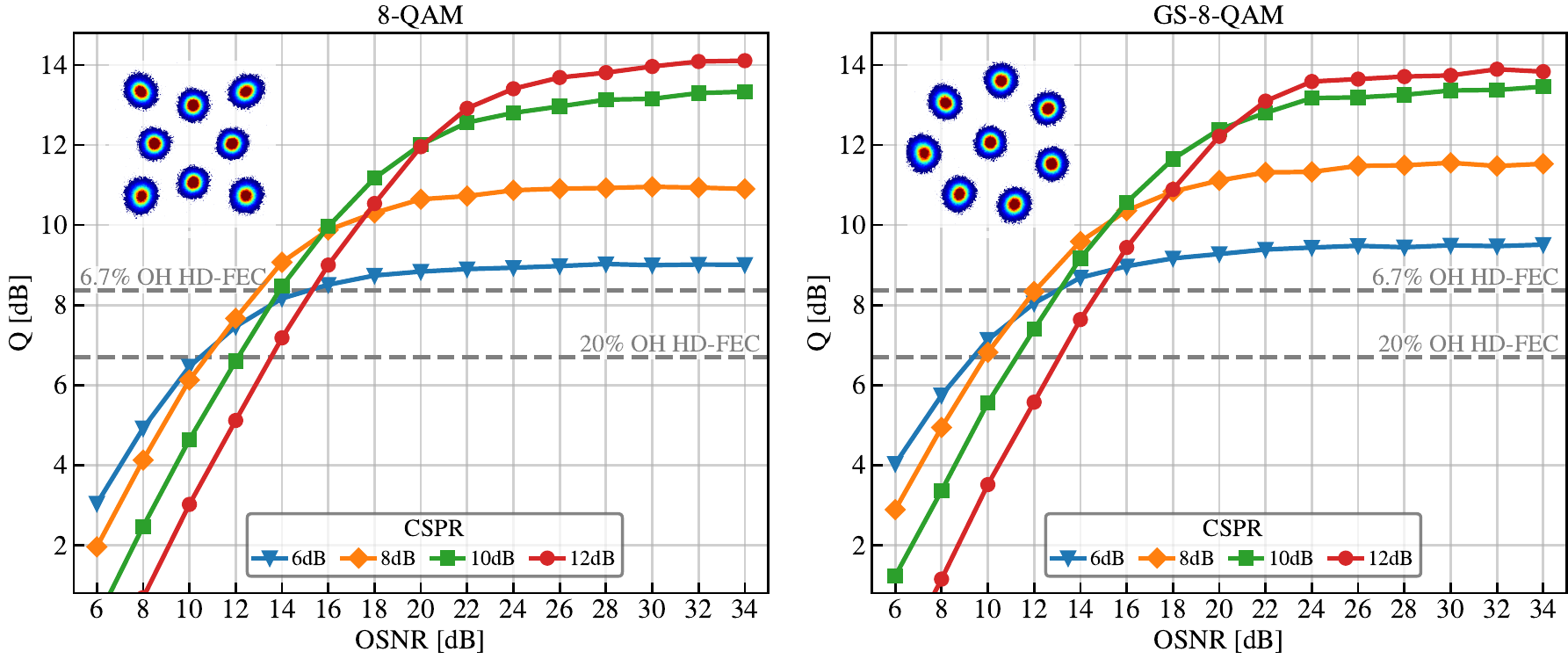}
\caption{Q-factor versus OSNR after transmission through \SI{91}{km} of field-deployed SSMF  for 8-QAM and GS-8-QAM for various CSPRs. Insets show constellations with highest Q-factor without noise loading  with \SI{12}{dB} \gls{CSPR}.}
\label{fig:osnr2} 
\end{figure*}

\cref{fig:osnr2} depicts the \gls{OSNR} performance for both conventional and geometrically-shaped \qam{8}. The \gls{GS} format outperforms its conventional counterpart and reaches the 6.7\% overhead threshold at \SI{12.0}{dB} versus \SI{13.0}{dB} for the conventional format. Similarly, at the 20\% threshold, \qam{GS-8} outperforms \qam{8} by \SI{1.0}{dB} and reaches the threshold at \SI{9.4}{dB} compared to \SI{10.4}{dB} for \qam{8}.

\begin{figure}
    \centering
    \begin{subfigure}{0.49\columnwidth}
		\includegraphics[width=\linewidth]{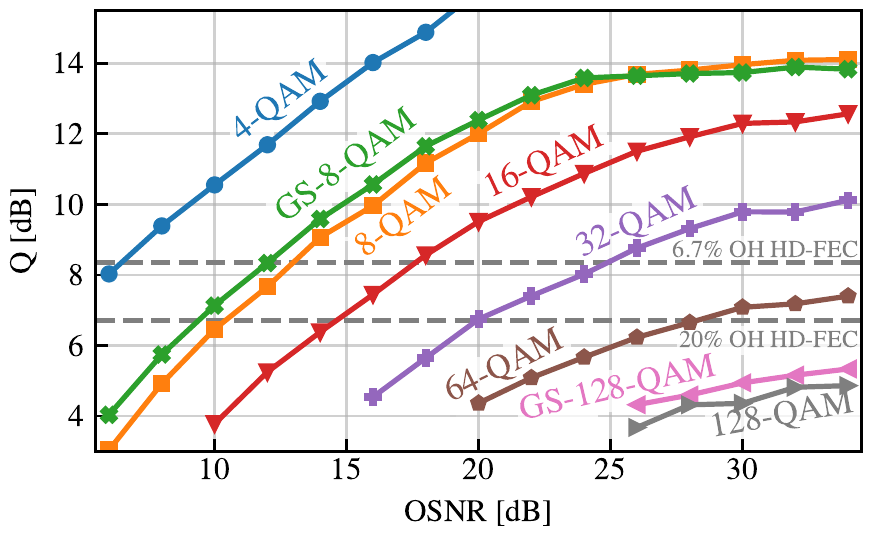}
		\caption{Q-factor versus OSNR for all evaluated modulation formats at optimized CSPR.}
		\label{fig:everythingplot} 
    \end{subfigure}%
    \hfill%
    \begin{subfigure}{0.49\columnwidth}
		\includegraphics[width=\linewidth]{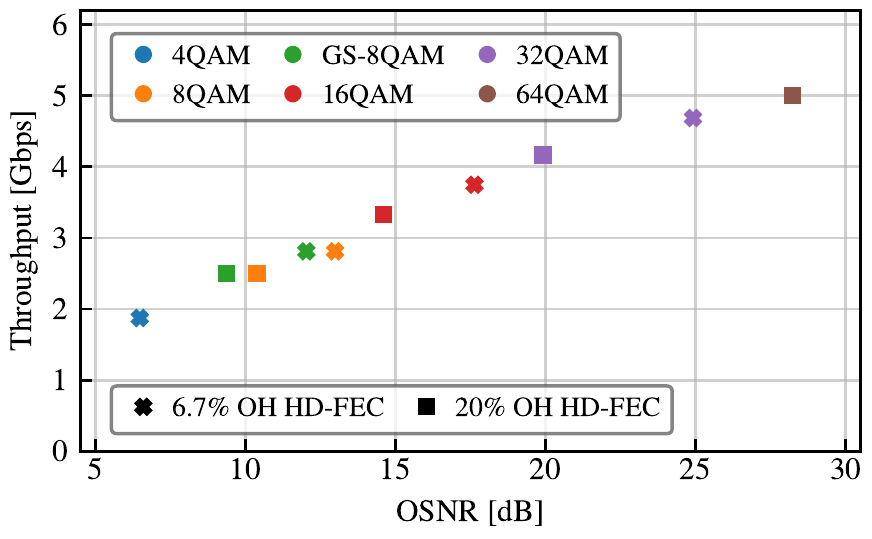}
		\caption{Net throughput as a function of OSNR for the evaluated modulation formats and HD-FEC thresholds.}
		\label{fig:throughputplot} 
    \end{subfigure}%
    \caption{Further analysis of all measurements.}
\end{figure}

\cref{fig:everythingplot} shows the Q-factor as a function of \gls{OSNR} for all evaluated modulation formats aggregated in one figure. For each \gls{OSNR}, the highest measured Q-factor for that modulation format is plotted, effectively tracing a line along the maximum Q-factor of each modulation format in \cref{fig:osnr0,fig:osnr1,fig:osnr2} and \qam{128- and GS-128}. As a result, \gls{CSPR} is optimized for each \gls{OSNR}. Using this optimization approach, \qam{4-, GS-8-, 8-, 16-, and 32} reach the 6.7\% \gls{HDFEC} threshold. \qam{64} only reaches the 20\% \gls{HDFEC} threshold, but \qam{128- and GS-128} unfortunately stay below the thresholds.

\cref{fig:throughputplot} estimates the net throughput of the transmission system at various \glspl{OSNR}. For each of the \gls{HDFEC} threshold crossings mentioned above, the net data rate after \gls{FEC} decoding at the \gls{OSNR} of the crossing is plotted. This figure reveals interesting choices for the system designer. If the system operates above an \gls{OSNR} of \SI{28.2}{dB}, \qam{64} combined with a 20\% overhead \gls{HDFEC} can be employed for a net data rate of \SI{5}{Gbps}. Alternatively, a lower complexity \gls{HDFEC} with an overhead of 6.7\% can be paired with \qam{32} for a net data rate of \SI{4.7}{Gbps}, requiring \SI{24.9}{dB} \gls{OSNR}. The multi-modulation format software-defined \gls{DSP} allows for efficient operation from \SI{6}{dB} up to \SI{28}{dB} \gls{OSNR}, flexibly switching modulation format depending on \gls{OSNR}.

\section{Discussion}
\label{sec:discussion}

\begin{figure}
    \centering
    \begin{subfigure}{0.25\columnwidth}
        \centering
        \includegraphics[width=0.8\textwidth]{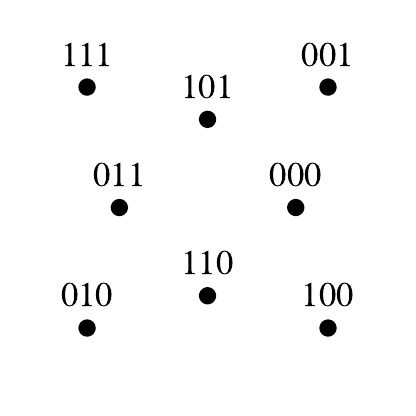}
        \caption{Conventional 8-QAM}
        \label{fig:8QAM}
    \end{subfigure}%
    \hfill%
    \begin{subfigure}{0.25\columnwidth}
        \centering
        \includegraphics[width=0.8\textwidth]{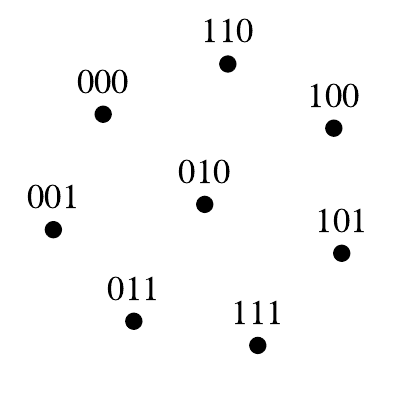}
        \caption{GS-8-QAM}
        \label{fig:GS8QAM}
    \end{subfigure}%
    \hfill%
    \begin{subfigure}{0.25\columnwidth}
        \centering
        \includegraphics[width=0.8\textwidth]{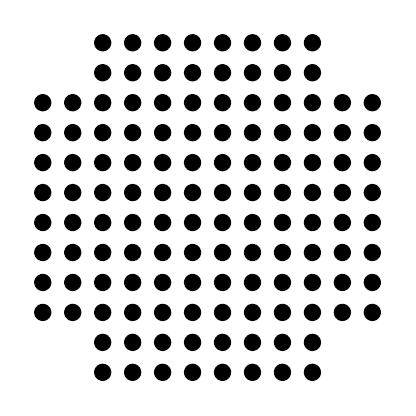}
        \caption{Conventional 128-QAM}
        \label{fig:128QAM}
    \end{subfigure}%
    \hfill%
    \begin{subfigure}{0.25\columnwidth}
        \centering
        \includegraphics[width=0.8\textwidth]{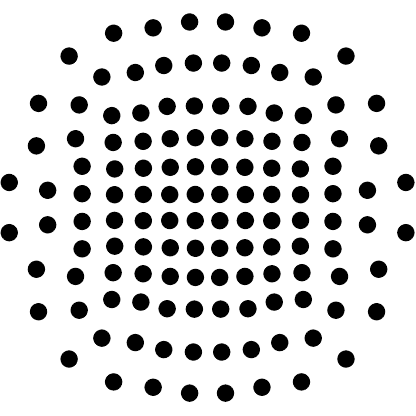}
        \caption{GS-128-QAM}
        \label{fig:GS128QAM}
    \end{subfigure}%
    \caption{Comparison between conventional and geometrically-shaped constellations. Drawn using the same scale with normalized power.}
    \label{fig:constellations}
\end{figure}

\cref{fig:constellations} shows the \gls{GS} formats used in this work compared with their conventional counterparts. Both \qam{GS-8} and \qam{GS-128} are optimized for the \gls{AWGN} channel at \gls{SNR} values of \SIlist{14;20}{dB}, respectively. The optimization uses an iterative approach similar to \cite{bin_chen_gs_2018}. The constellations are initialized using the conventional layout and optimized by iterating between adding perturbations in the form of Gaussian noise to a randomly chosen single point and swapping the binary labels of two randomly chosen constellation points until convergence is reached. After each iteration, the \gls{GMI} for the \gls{AWGN} channel is evaluated and if gains are found, the modified constellation is taken as the new baseline. \qam{GS-8}, see \cref{fig:GS8QAM}, resembles a circular \qam{8} with center constellation point \cite{Nolle:14}, but is different since it is not symmetric. Symmetries are added to \qam{GS-128} to aid convergence of the constellation design. The result of the optimization algorithm is to increase the Euclidean distance between constellation points that differ more than 1 bit.

\begin{figure*}[!t]
\includegraphics[width=\linewidth]{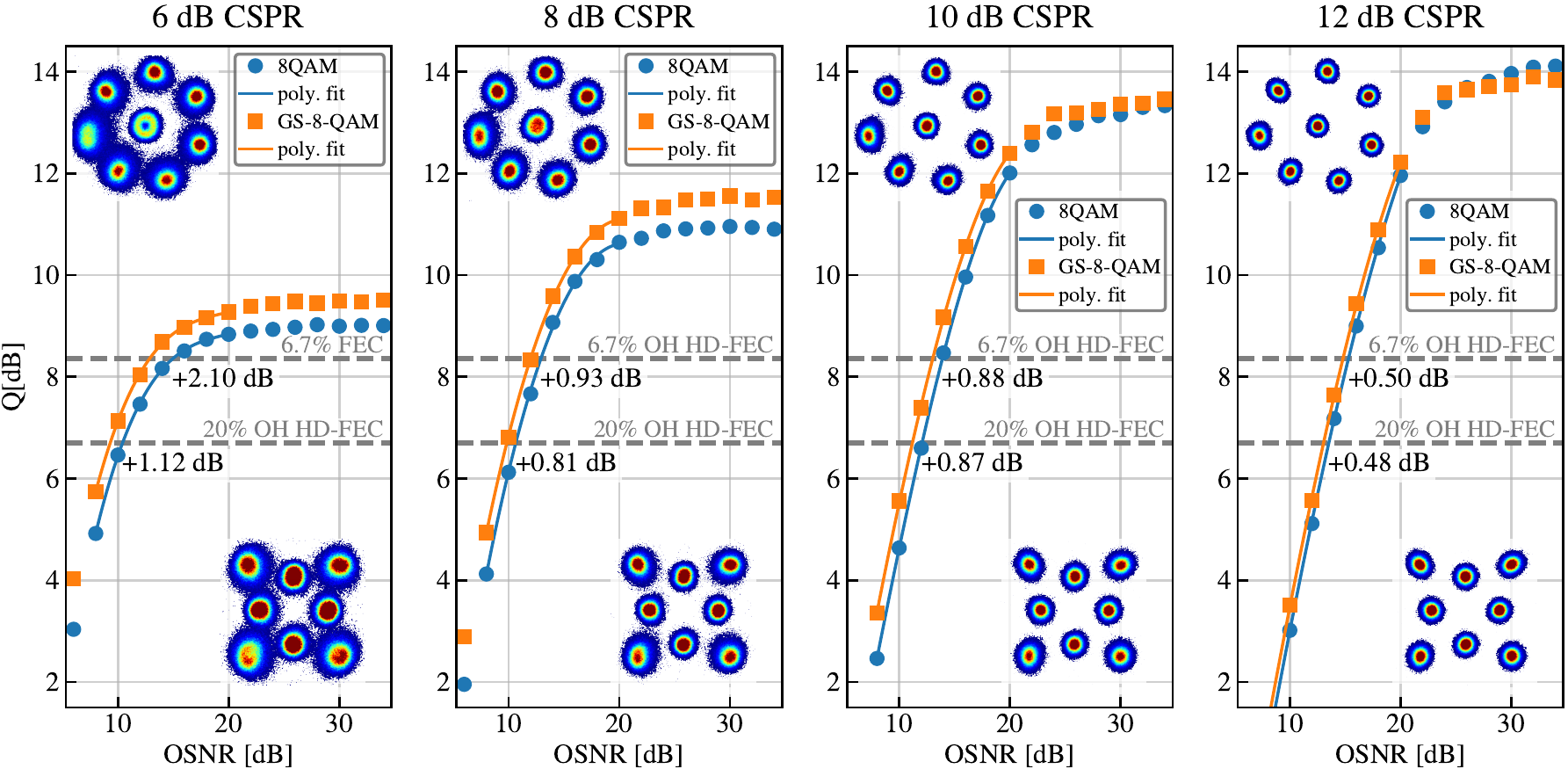}
\caption{Q-factor versus OSNR after transmission through \SI{91}{km} of field-deployed SSMF for 8-QAM and GS-8-QAM for various CSPRs. Values indicate OSNR gain of GS-8-QAM with respect to 8-QAM at the specified FEC threshold calculated using a polynomial fit. Insets show constellations with highest Q-factor at \SI{34}{dB} OSNR.}
\label{fig:GSgain} 
\end{figure*}

\Gls{GS} formats for \glspl{SNR} were created in steps of \SI{1}{dB} ranging from \SIrange{5}{14}{dB} and \SIrange{17}{23}{dB} for \qam{8} and \qam{128}, respectively, and were experimentally tested using the field-deployed link since there is no a priori knowledge of the \gls{AWGN} channel that best resembles the transmission scenario. Many of the \gls{GS} formats tested on the transmission link perform worse than their conventional counterparts, but \qam{GS-8} optimized for \SI{14}{dB} \gls{SNR} and \qam{GS-128} optimized for \SI{20}{dB} show significant experimental gains and are included in this work. The constellation design technique optimizes for \gls{AWGN}, which is a substantially different channel than the one employed in this work. The gain of \qam{GS-8} with respect to \qam{8} in the \gls{AWGN} channel is \SI{0.5}{dB} at both \gls{FEC} thresholds. \cref{fig:GSgain} shows the influence of \gls{KK} reconstruction errors on \gls{OSNR} for various \glspl{CSPR}. The insets shows that the outer points of a constellation are predominantly distorted. The employed \qam{GS-8} is more resilient against these distortions than \qam{8}, enabling operation at lower \gls{CSPR}. Therefore, it is expected that more advanced optimization techniques taking into account the channel including \gls{CSPR} and \gls{KK} reconstruction errors will produce better performing modulation formats. However, determining the best \gls{GS} modulation format for the transmission scenario was not the goal of this investigation. These suboptimal constellations are included as a proof-of-concept for using \gls{GS} formats in the flexible multi-modulation format receiver, to demonstrate that the receiver supports non-symmetrical modulation formats, and to demonstrate the use of \qam{GS} in \gls{KK} coherent detection.

In \cite{gpuJLT}, we note that the error floor or Q-factor ceiling of high-cardinality formats is most likely due to low-pass filtering at the receiver causing \gls{KK} reconstruction errors. Here, we use a photodiode with higher bandwidth, enabling successful processing of \qam{64} signals reaching the 20\% overhead \gls{HDFEC} Q-factor threshold. Further gains could be made by employing a higher bandwidth \gls{ADC} as well. Note that the \gls{KK} reconstruction errors are caused by filtering effects between direct detection at the photodiode and conversion in the \gls{ADC} \cite{chenKKDCI}.  Alternatively, one could implement a second static equalizer before the \gls{KK} algorithm to counter these filtering effects \cite{chenKKFE}. The current implementation almost fully utilizes the \gls{GPU} processing capabilities, but further optimization of the implementation can free up resources for such an additional filtering step.

The main factor limiting baud and data rates in this work is the \gls{ADC}. The off-the-shelf commercial 12-bit \SI{4}{GS/s} \gls{ADC} has a \SI{3}{dB} bandwidth of \SI{1}{GHz}, limiting the baud rate to \SI{1}{GBaud}. The \gls{ADC} uses an eight-lane PCIe Gen 3 interface to communicate with the \gls{GPU}, which limits the sampling rate of the \gls{ADC}. Future \glspl{ADC}, employing PCIe Gen 4 or additional lanes, are expected to offer greater sampling rates and bandwidth. Proprietary interfaces such as NVIDIA NVLink (at the time of writing is not available in off-the-shelf \glspl{ADC}) can support an order of magnitude higher throughput from \gls{ADC} to \gls{GPU} and may enable greater increases in sampling rates.

\section{Conclusion}
\label{sec:conclusion}

A commercial off-the-shelf \gls{GPU} is used for real-time digital signal processing of minimum-phase \qam{4-, 8-, 16-, 32-, 64-, and 128} as well as geometrically-shaped \qam{8} and \qam{128} signals detected with a \gls{KK} coherent receiver \cite{mecozzi_kramers_2016}. This real-time, flexible, multi-modulation format receiver is experimentally validated using a field-deployed link between two Tokyo locations. A net data rate of \SI{5}{Gbps} is demonstrated using \SI{1}{GBaud} \qam{64}. This shows the potential of \glspl{GPU} for software-defined signal processing functionality in optical communication systems.

\begin{backmatter}

\bmsection{Funding}
This work was partly supported by the Dutch NWO Gravitation Program on Research Center for Integrated Nanophotonics under Grant GA~024.002.033. The work of A. Alvarado and S. Goossens has received funding from the European Research Council (ERC) under the European Union’s Horizon 2020 research and innovation programme (grant agreement No. 757791).

\bmsection{Data availability}
Data underlying the results presented in this paper are not publicly available at this time but may be obtained from the authors upon reasonable request.

\bmsection{Disclosures}
The authors declare no conflicts of interest
\end{backmatter}


\printbibliography[notcategory=ignore]

@article{Luis:20,
    author = {Ruben {S. Lu\'is} and Georg Rademacher and Benjamin J. Puttnam and Cristian Antonelli and Satoshi Shinada and Hideaki Furukawa},
    journal = {Opt. Express},
    number = {3},
    pages = {4067--4075},
    publisher = {OSA},
    title = {{Simple method for optimizing the DC bias of Kramers-Kronig receivers based on AC-coupled photodetectors}},
    volume = {28},
    month = {2},
    year = {2020},
    doi = {10.1364/OE.383369},
}

@article{winzer_scaling_2017,
	title = {{From Scaling Disparities to Integrated Parallelism: {A} Decathlon for a Decade}},
	volume = {35},
	number = {5},
	journal = {J. of Lightw. Technol.},
	author = {Winzer, P. J. and Neilson, D. T.},
	month = {3},
	year = {2017},
	pages = {1099--1115},
	doi={10.1109/JLT.2017.2662082},
}

@ARTICLE{suzuki_phy_2020,
  author={T. {Suzuki} and S. {Kim} and J. {Kani} and J. {Terada}},
  journal={Journal of Lightwave Technology}, 
  title={{Demonstration of Fully Softwarized 10G-EPON PHY Processing on a General-Purpose Server for Flexible Access Systems}}, 
  year={2020},
  volume={38},
  number={4},
  pages={777-783},
  doi={10.1109/JLT.2019.2948333}}

@article{suzuki_phy_2019,
	title = {Software {Implementation} of {10G}-{EPON} {Upstream} {Physical}-{Layer} {Processing} for {Flexible} {Access} {Systems}},
	volume = {37},
	number = {6},
	journal = {J. of Lightw. Technol.},
	author = {Suzuki, Takahiro and Kim, Sang Yuep and Kani, Jun Ichi and Terada, Jun},
	month = {3},
	year = {2019},
	pages = {1631--1637},
	doi={10.1109/ACCESS.2019.2904083},
}

@ARTICLE{suzuki_phy_2018,
  author={T. {Suzuki} and S. {Kim} and J. {Kani} and A. {Otaka} and T. {Hanawa}},
  journal={Journal of Lightwave Technology}, 
  title={{10-Gb/s Software Implementation of Burst-Frame Synchronization Using Array-Access Bitshift and Dual-Stage Detection for Flexible Access Systems}}, 
  year={2018},
  volume={36},
  number={23},
  pages={5656-5662},
  doi={10.1109/JLT.2018.2870912}}

@article{suzuki_real-time_2020,
	title = {Real-{Time} {Implementation} of {Coherent} {Receiver} {DSP} {Adopting} {Stream} {Split} {Assignment} on {GPU} for {Flexible} {Optical} {Access} {Systems}},
	volume = {38},
	number = {3},
	journal = {J. of Lightw. Technol.},
	author = {Suzuki, T and Kim, S and Kani, J and Terada, J},
	month = {2},
	year = {2020},
	pages = {668--675},
	doi={10.1109/JLT.2019.2950155},
}

@article{mecozzi_kramers_2016,
	title = {Kramers {Kronig} coherent receiver},
	volume = {3},
	shorttitle = {Kramers?},
	doi = {10.1364/OPTICA.3.001220},
	number = {11},
	journal = {Optica},
	author = {Mecozzi, Antonio and Antonelli, Cristian and Shtaif, Mark},
	month = {11},
	year = {2016},
	pages = {1220}
}

@article{sun_summarizing_2019,
	title = {Summarizing {CPU} and {GPU} {Design} {Trends} with {Product} {Data}},
	journal = {arXiv:1911.11313 [cs]},
	author = {Sun, Yifan and Agostini, Nicolas Bohm and Dong, Shi and Kaeli, David},
	month = nov,
	year = {2019},
}

@ARTICLE{Silva_WidelyLinear,  
author={E. P. {da Silva} and D. {Zibar}},  
journal={Journal of Lightwave Technology},   
title={{Widely Linear Equalization for IQ Imbalance and Skew Compensation in Optical Coherent Receivers}},   
year={2016},  
volume={34},  
number={15},  
pages={3577-3586},
doi={10.1109/JLT.2016.2577716}}

@INPROCEEDINGS{Li_realtime_LDPC,
  author={R. {Li} and J. {Zhou} and Y. {Dou} and S. {Guo} and D. {Zou} and S. {Wang}},
  booktitle={2013 IEEE 14th Workshop on Signal Processing Advances in Wireless Communications (SPAWC)}, 
  title={{A multi-standard efficient column-layered LDPC decoder for Software Defined Radio on GPUs}}, 
  year={2013},
  volume={},
  number={},
  pages={724-728},
  doi={10.1109/SPAWC.2013.6612145}}

@ARTICLE{suzuki_fec,
  author={T. {Suzuki} and S. {Kim} and J. {Kani} and T. {Hanawa} and K. {Suzuki} and A. {Otaka}},
  journal={Journal of Lightwave Technology}, 
  title={{Demonstration of 10-Gbps Real-Time Reed–Solomon Decoding Using GPU Direct Transfer and Kernel Scheduling for Flexible Access Systems}}, 
  year={2018},
  volume={36},
  number={10},
  pages={1875-1881},
  doi={10.1109/JLT.2018.2793938}}

@INPROCEEDINGS{Randel_FPGA_2015,
  author={S. {Randel} and S. {Corteselli} and D. {Badini} and D. {Pilori} and S. {Caelles} and S. {Chandrasekhar} and J. {Gripp} and H. {Chen} and N. K. {Fontaine} and R. {Ryf} and P. J. {Winzer}},
  booktitle={2015 IEEE Photonics Conference (IPC)}, 
  title={{First real-time coherent MIMO-DSP for six coupled mode transmission}}, 
  year={2015},
  volume={},
  number={},
  doi={10.1109/IPCon.2015.7323761}}

@inproceedings{Beppu_FPGA_2020,
author = {S. Beppu and K. Igarashi and H. Mukai and M. Kikuta and M. Shigihara and D. Soma and T. Tsuritani and I. Morita},
booktitle = {Optical Fiber Communication Conference (OFC) 2020},
journal = {Optical Fiber Communication Conference (OFC) 2020},
pages = {Th3H.2},
publisher = {Optical Society of America},
title = {{Real-time strongly-coupled 4-core fiber transmission}},
year = {2020},
doi = {10.1364/OFC.2020.Th3H.2},
}

@article{Beppu_2_FPGA_2020,
author = {Shohei Beppu and Koji Igarashi and Masahiro Kikuta and Daiki Soma and Tomoyuki Nagai and Yasuo Saito and Hidenori Takahashi and Takehiro Tsuritani and Itsuro Morita and Masatoshi Suzuki},
journal = {Opt. Express},
number = {13},
pages = {19655--19668},
publisher = {OSA},
title = {{Weakly coupled 10-mode-division multiplexed transmission over 48-km few-mode fibers with real-time coherent MIMO receivers}},
volume = {28},
month = {6},
year = {2020},
doi = {10.1364/OE.395415},
}

@article{ECOC_realtimeGPU,
  title={{Real-time, Software-Defined, GPU-Based Receiver Field Trial}},
  author={van der Heide, Sjoerd P and Luis, Ruben S and Puttnam, Benjamin J and Rademacher, Georg and Koonen, Ton and Shinada, Satoshi and Awaji, Yoshinari and Okonkwo, Chigo and Furukawa, Hideaki},
  journal={ECOC We1E5},
  year={2020},
  doi={10.1109/ECOC48923.2020.9333244},
}

@article{gpuJLT,
author = {Sjoerd P van der Heide and Ruben S. Luis and Benjamin J. Puttnam and Georg Rademacher and Ton Koonen and Satoshi Shinada and Yohinari Awaji and Hideaki Furukawa and Chigo Okonkwo},
journal = {J. Lightwave Technol.},
number = {8},
pages = {2358--2367},
publisher = {OSA},
title = {{Field Trial of a Flexible Real-Time Software-Defined GPU-Based Optical Receiver}},
volume = {39},
month = {4},
year = {2021},
doi = {10.1109/JLT.2021.3050304},
}

@inproceedings{OFC_realtime10k,
author = {van der Heide, S P and Luis, R S and Puttnam, B J and Rademacher, G and Koonen, T and Shinada, S and Awaji, Y and Furukawa, H and Okonkwo, C},
booktitle = {2021 Optical Fiber Communications Conference and Exhibition (OFC)},
pages = {1--3},
title = {{10,000 km Straight-line Transmission using a Real-time Software-defined GPU-Based Receiver}},
year = {2021},
doi = {10.1364/OFC.2021.W1I.3},
}

@INPROCEEDINGS{bin_chen_gs_2018,
  author={Chen, Bin and Okonkwo, Chigo and Lavery, Domaniç and Alvarado, Alex},
  booktitle={2018 20th International Conference on Transparent Optical Networks (ICTON)},
  title={Geometrically-shaped 64-point Constellations via Achievable Information Rates}, 
  year={2018},
  volume={},
  number={},
  pages={1-4},
  doi={10.1109/ICTON.2018.8473932}}

@ARTICLE{chenKKDCI,  author={Chen, Xi and Antonelli, Cristian and Chandrasekhar, Sethumadhavan and Raybon, Gregory and Mecozzi, Antonio and Shtaif, Mark and Winzer, Peter},  journal={Journal of Lightwave Technology},   title={Kramers–Kronig Receivers for 100-km Datacenter Interconnects},   year={2018},  volume={36},  number={1},  pages={79-89},  doi={10.1109/JLT.2018.2793460}}

@INPROCEEDINGS{chenKKFE,  author={Chen, X. and Chandrasekhar, S. and Olsson, S. and Adamiecki, A. and Winzer, P.},  booktitle={2018 European Conference on Optical Communication (ECOC)},   title={{Impact of O/E Front-End Frequency Response on Kramers-Kronig Receivers and its Compensation}},   year={2018},  volume={},  number={},  pages={1-3},  doi={10.1109/ECOC.2018.8535239}}

@inproceedings{Nolle:14,
author = {N{\"{o}}lle, Markus and Frey, Felix and Elschner, Robert and Schmidt-Langhorst, Carsten and Napoli, Antonio and Schubert, Colja},
booktitle = {Optical Fiber Communication Conference},
doi = {10.1364/OFC.2014.W3B.2},
pages = {W3B.2},
publisher = {Optical Society of America},
title = {{Performance Comparison of Different 8QAM Constellations for the Use in Flexible Optical Networks}},
year = {2014}
}

@inproceedings{7pctFEC,
author = {Miyata, Y and Kubo, K and Onohara, K and Matsumoto, W and Yoshida, H and Mizuochi, T},
booktitle = {OFC/NFOEC},
pages = {1--3},
title = {{UEP-BCH product code based hard-decision FEC for 100 Gb/s optical transport networks}},
year = {2012},
doi={10.1364/NFOEC.2012.JW2A.7},
}

@inbook{20pctFEC2,
address = {Cham},
author = {i Amat, Alexandre and Schmalen, Laurent},
booktitle = {Springer Handbook of Optical Networks},
chapter = {7},
doi = {10.1007/978-3-030-16250-4_7},
isbn = {978-3-030-16250-4},
pages = {177--257},
publisher = {Springer International Publishing},
title = {{Forward Error Correction for Optical Transponders}},
year = {2020},
note = {Table 7.5}
}

@inproceedings{20pctFEC1,
author = {Scholten, Michael and Coe, Tim and Dillard, John and Chang, Frank},
booktitle = {2009 European Conference on Optical Communication (ECOC)},
pages = {1--12},
title = {{Enhanced FEC for 40G/100G}},
year = {2009}
}






\end{document}